\documentclass[english]{article}
\usepackage[T1]{fontenc}
\usepackage[latin9]{inputenc}
\usepackage{geometry}
\usepackage{amsmath}

\makeatletter
\newcommand{\lyxaddress}[1]{
	\par {\raggedright #1
	\vspace{1.4em}
	\noindent\par}
}

\@ifundefined{date}{}{\date{}}
\makeatother

\usepackage{babel}
\begin{document}
\title{Comment on ``Lanczos potential of Weyl field: \\
interpretations and applications''}
\author{Ronald J. Riegert\thanks{Email: rriegert@ucsd.edu}}
\maketitle

\lyxaddress{\begin{center}
Las Cruces, New Mexico USA
\par\end{center}}

In the recently published review article \cite{key-1} Vishwakarma
considers, among other topics, the question of how the Lanczos potential
$L_{\mu\nu\sigma}$ in a four-dimensional Riemannian spacetime behaves
under a conformal transformation. In attempting to formulate a consistent
conformally-covariant treatment of the potential, the author is led
to impose a complicated algebraic constraint among the components
of $L_{\mu\nu\sigma}$. However, we show below that Vishwakarma's
proposed constraint is incompatible with the existence of a Lanczos
potential in a large class of spacetimes and therefore must be rejected.
We briefly recap the author's argument (denoting equation numbers
that appear in \cite{key-1} with the prefix ``V'') and follow with
our critique\footnote{\noindent Tensor calculations were performed in Wolfram \emph{Mathematica}
using the \emph{xAct} suite of packages for computer algebra\cite{key-2}.}.

Vishwakarma begins with the Weyl-Lanczos equation, expressing the
Weyl conformal curvature tensor $C_{\mu\nu\sigma\rho}$ as a linear
combination of first-derivatives of the potential $L_{\mu\nu\sigma}$:
\begin{align*}
C_{\mu\nu\sigma\rho} & =L_{\mu\nu\sigma;\rho}+L_{\sigma\rho\mu;\nu}-L_{\mu\nu\rho;\sigma}-L_{\sigma\rho\nu;\mu}+g_{\nu\sigma}L_{\mu\rho}+g_{\mu\rho}L_{\nu\sigma}\\
 & \qquad-g_{\nu\rho}L_{\mu\sigma}-g_{\mu\sigma}L_{\nu\rho}+\frac{2}{3}L^{\lambda\kappa}{}_{\lambda;\kappa}\left(g_{\mu\sigma}g_{\nu\rho}-g_{\nu\sigma}g_{\mu\rho}\right)\tag{V4}\\
 & \equiv W_{\mu\nu\sigma\rho}
\end{align*}
where $L_{\mu\nu}=L_{\nu\mu}\equiv\tfrac{1}{2}\left(L_{\mu}{}^{\kappa}{}_{\nu}{}_{;\kappa}+L_{\nu}{}^{\kappa}{}_{\mu}{}_{;\kappa}-L_{\mu}{}^{\kappa}{}_{\kappa}{}_{;\nu}-L_{\nu}{}^{\kappa}{}_{\kappa}{}_{;\mu}\right)$.
Next, the author considers how eq.(V4) changes under the local conformal
rescaling $g_{\mu\nu}\rightarrow\tilde{g}_{\mu\nu}=\Omega^{2}(x)g_{\mu\nu},L_{\mu\nu\sigma}\rightarrow\tilde{L}_{\mu\nu\sigma}=\Omega^{s}(x)L_{\mu\nu\sigma}$
(where \textit{s} is to be determined) and finds that the right side
transforms according to:
\begin{align*}
\tilde{W}^{\mu}{}_{\nu\sigma\rho} & =\Omega^{(s-2)}W^{\mu}{}_{\nu\sigma\rho}\\
 & \qquad+\frac{(s-3)}{2}\Omega^{(s-3)}g^{\mu\alpha}\left[2\left\{ L_{\alpha\nu\sigma}\Omega_{,\rho}+L_{\nu\alpha\rho}\Omega_{,\sigma}+L_{\sigma\rho\alpha}\Omega_{,\nu}+L_{\rho\sigma\nu}\Omega_{,\alpha}\right\} \right.\\
 & \qquad+\left\{ g_{\alpha\rho}\left(L_{\nu}{}^{\kappa}{}_{\sigma}+L_{\sigma}{}^{\kappa}{}_{\nu}\right)+g_{\nu\sigma}\left(L_{\alpha}{}^{\kappa}{}_{\rho}+L_{\rho}{}^{\kappa}{}_{\alpha}\right)-g_{\alpha\sigma}\left(L_{\nu}{}^{\kappa}{}_{\rho}+L_{\rho}{}^{\kappa}{}_{\nu}\right)\tag{V31}\right.\\
 & \qquad\left.-g_{\nu\rho}\left(L_{\alpha}{}^{\kappa}{}_{\sigma}+L_{\sigma}{}^{\kappa}{}_{\alpha}\right)\right\} \Omega_{,\kappa}+\left\{ \left(g_{\nu\rho}L_{\sigma}{}^{\kappa}{}_{\kappa}-g_{\nu\sigma}L_{\rho}{}^{\kappa}{}_{\kappa}\right)\Omega_{,\alpha}+\left(g_{\alpha\sigma}L_{\rho}{}^{\kappa}{}_{\kappa}-g_{\alpha\rho}L_{\sigma}{}^{\kappa}{}_{\kappa}\right)\Omega_{,\nu}\right.\\
 & \qquad+\left(g_{\rho\nu}L_{\alpha}{}^{\kappa}{}_{\kappa}-g_{\rho\alpha}L_{\nu}{}^{\kappa}{}_{\kappa}\right)\Omega_{,\sigma}+\left(g_{\sigma\alpha}L_{\nu}{}^{\kappa}{}_{\kappa}-g_{\sigma\nu}L_{\alpha}{}^{\kappa}{}_{\kappa}\right)\Omega_{,\rho}\}-\frac{4}{3}L^{\lambda\kappa}{}_{\kappa}\Omega_{,\lambda}\left(g_{\alpha\sigma}g_{\nu\rho}-g_{\nu\sigma}g_{\alpha\rho}\right)]
\end{align*}
Demanding scale invariance, i.e., $\tilde{W}^{\mu}{}_{\nu\sigma\rho}=W^{\mu}{}_{\nu\sigma\rho}$,
Vishwakarma then sets $s=2$ and discards the additive term proportional
to $(s-3)$ by proposing, without proof, that the potential $L_{\mu\nu\sigma}$
can always be chosen to satisfy the intricate ``symmetry'' condition:
\begin{align*}
\left[2\left\{ L_{\alpha\nu\sigma}\Omega_{,\rho}+L_{\nu\alpha\rho}\Omega_{,\sigma}+L_{\sigma\rho\alpha}\Omega_{,\nu}+L_{\rho\sigma\nu}\Omega_{,\alpha}\right\} \right.\\
+\left\{ g_{\alpha\rho}\left(L_{\nu}{}^{\kappa}{}_{\sigma}+L_{\sigma}{}^{\kappa}{}_{\nu}\right)+g_{\nu\sigma}\left(L_{\alpha}{}^{\kappa}{}_{\rho}+L_{\rho}{}^{\kappa}{}_{\alpha}\right)-g_{\alpha\sigma}\left(L_{\nu}{}^{\kappa}{}_{\rho}+L_{\rho}{}^{\kappa}{}_{\nu}\right)\right. & \tag{V32}\\
\left.-g_{\nu\rho}\left(L_{\alpha}{}^{\kappa}{}_{\sigma}+L_{\sigma}{}^{\kappa}{}_{\alpha}\right)\right\} \Omega_{,\kappa}+\left\{ \left(g_{\nu\rho}L_{\sigma}{}^{\kappa}{}_{\kappa}-g_{\nu\sigma}L_{\rho}{}^{\kappa}{}_{\kappa}\right)\Omega_{,\alpha}+\left(g_{\alpha\sigma}L_{\rho}{}^{\kappa}{}_{\kappa}-g_{\alpha\rho}L_{\sigma}{}^{\kappa}{}_{\kappa}\right)\Omega_{,\nu}\right.\\
+\left(g_{\rho\nu}L_{\alpha}{}^{\kappa}{}_{\kappa}-g_{\rho\alpha}L_{\nu}{}^{\kappa}{}_{\kappa}\right)\Omega_{,\sigma}+\left(g_{\sigma\alpha}L_{\nu}{}^{\kappa}{}_{\kappa}-g_{\sigma\nu}L_{\alpha}{}^{\kappa}{}_{\kappa}\right)\Omega_{,\rho}\}-\frac{4}{3}L^{\lambda\kappa}{}_{\kappa}\Omega_{,\lambda}\left(g_{\alpha\sigma}g_{\nu\rho}-g_{\nu\sigma}g_{\alpha\rho}\right)] & =0
\end{align*}
for arbitrary $\Omega$. 

But this is easily falsified without any need to compute Lanczos potentials
and test them in the cumbersome rank-4 condition (V32). We instead
form a much simpler necessary condition of rank-0 by first contracting
eq.(V32) with $C^{\alpha\nu\sigma\rho}$ to get $8\thinspace C^{\alpha\nu\sigma\rho}L_{\alpha\nu\sigma}\Omega_{,\rho}=0$.
Because the function $\Omega(x)$ is arbitrary, this requires that
the coefficient vector $Q^{\rho}\equiv C^{\alpha\nu\sigma\rho}L_{\alpha\nu\sigma}=0$
throughout all spacetime. And since $Q^{\rho}$ vanishes everywhere,
then so must its covariant derivative $Q^{\rho}{}_{;\tau}$ and, more
specifically, its covariant divergence $Q^{\rho}{}_{;\rho}$. The
scalar condition $Q^{\rho}{}_{;\rho}=0$ is thus necessary (but of
course, not sufficient) for the validity of eq.(V32). To check whether
this divergence does indeed vanish everywhere, we apply the product
rule:
\[
Q^{\rho}{}_{;\rho}=C^{\alpha\nu\sigma\rho}{}_{;\rho}L_{\alpha\nu\sigma}+C^{\alpha\nu\sigma\rho}L_{\alpha\nu\sigma;\rho}
\]
and then use the once-contracted Bianchi identity:
\[
C^{\alpha\nu\sigma\rho}{}_{;\rho}\equiv\tfrac{1}{2}\left(R^{\alpha\sigma;\nu}-R^{\nu\sigma;\alpha}\right)-\tfrac{1}{12}\left(g^{\alpha\sigma}R^{;\nu}-g^{\nu\sigma}R^{;\alpha}\right)
\]
along with the ``scalarized'' Weyl-Lanczos equation obtained by
the contraction of eq.(V4) with the Weyl tensor:
\[
C^{\alpha\nu\sigma\rho}C_{\alpha\nu\sigma\rho}=4\,C^{\alpha\nu\sigma\rho}L_{\alpha\nu\sigma;\rho}
\]
to rewrite it as:
\[
Q^{\rho}{}_{;\rho}=\left(R^{\alpha\sigma;\nu}-\tfrac{1}{6}\thinspace g^{\alpha\sigma}R^{;\nu}\right)L_{\alpha\nu\sigma}+\tfrac{1}{4}C^{\alpha\nu\sigma\rho}C_{\alpha\nu\sigma\rho}
\]
Specializing now to Ricci-flat (``Rf'') $R^{\alpha\sigma}=0$ spacetimes,
only the term quadratic in the Weyl-tensor survives on the right side:
\[
Q_{(Rf);\rho}^{\rho}=\tfrac{1}{4}C^{\alpha\nu\sigma\rho}C_{\alpha\nu\sigma\rho}=\tfrac{1}{4}K
\]
where $K$ is the Kretschmann scalar. For example, in the Ricci-flat
Schwarzschild (``Sc'') spacetime with line element:
\[
ds^{2}=\left(1-\frac{2GM}{r}\right)dt^{2}-\left(1-\frac{2GM}{r}\right)^{-1}dr^{2}-r^{2}\left(d\theta^{2}+\sin^{2}\theta\,d\varphi^{2}\right)
\]
this divergence evaluates to $Q_{(Sc);\rho}^{\rho}=\tfrac{1}{4}K_{(Sc)}=\frac{12\thinspace G^{2}M^{2}}{r^{6}}$,
which is clearly \textit{not} zero for any finite radius \textit{r}
or at any time \textit{t}. In other words, the Lanczos potential $L_{(Sc)\alpha\nu\sigma}$
in a Schwarzschild spacetime with mass $M\neq0$ violates eq.(V32).
Indeed, the potentials in \textit{all} generic Ricci-flat spacetimes
with $K\neq0$ fail condition (V32), and we speculate that those in
most Ricci-curved spacetimes likely fail as well. 

So while there might exist particular spacetimes with Lanczos potentials
that do happen to obey eq.(V32), we conclude that Vishwakarma's scheme
to implement conformal-covariance for the Weyl-Lanczos equation by
setting $s=2$ is not viable in the general case. Instead, we propose
to accept that the Lanczos potential $L_{\mu\nu\sigma}$ most naturally
scales with the weight $s=3$ and investigate whether the Weyl-Lanczos
equation itself can be altered to realize uniform overall scaling
with this weighting. This possibility will be examined elsewhere.

\end{document}